%% file: main.tex
\documentclass[a4paper,journal]{IEEEtran} 
\IEEEoverridecommandlockouts
\usepackage[left=0.625in,right=0.625in,top=0.7in,bottom=1.1in]{geometry}
\usepackage[dvipsnames]{xcolor}
\usepackage{cite}
\usepackage{amsmath,amssymb,amsfonts}
\usepackage{graphicx}
\usepackage{textcomp}
\usepackage{xcolor}
\usepackage{float}
\usepackage{cite}
\usepackage{placeins}
\usepackage{textcomp}
\usepackage{multirow}
\usepackage{xcolor}
\usepackage{wrapfig}
\usepackage{algpseudocode}
\usepackage{algorithm}
\usepackage{caption}
\usepackage{subcaption}
\usepackage{comment}

\usepackage[noabbrev]{cleveref}

\usepackage{tabularx, multirow,multicol,colortbl,array}
\usepackage{graphicx,color, array}
\usepackage[dvipsnames]{xcolor}
\usepackage{float}
\usepackage{comment,psfrag}
\usepackage{enumitem}
\usepackage{tikz,tikzscale,bigints,pgfplots}
\pgfplotsset{compat=1.18}
\usetikzlibrary{positioning, arrows.meta, colorbrewer, calc, fit, matrix}
\usepackage{multirow}

\usepackage{import}
\usepackage{pgfplots}
\usetikzlibrary{calc}
\usetikzlibrary{spy}
\usetikzlibrary{matrix}
\usetikzlibrary{backgrounds}
\pgfplotsset{compat = newest}

\usetikzlibrary{shapes.misc, positioning}
\usetikzlibrary{shadows,arrows,shapes,calc,positioning,decorations.pathreplacing,fit}
\def\BibTeX{{\rm B\kern-.05em{\sc i\kern-.025em b}\kern-.08em
		T\kern-.1667em\lower.7ex\hbox{E}\kern-.125emX}}

\makeatletter
\def\endthebibliography{%
  \def\@noitemerr{\@latex@warning{Empty `thebibliography' environment}}%
  \endlist
}
\makeatother

\begin{document}

\pgfplotsset{
    standard/.style={
    axis line style = thick,
    grid = both,
    }
}

\pagestyle{empty}

\title{Modified Baum-Welch Algorithm for Joint Blind Channel Estimation and Turbo Equalization

}

\author{
    \IEEEauthorblockN{ Chin-Hung Chen$^{\star}$}
    , \IEEEauthorblockN{Boris Karanov$^{\dagger}$}%
    , \IEEEauthorblockN{Ivana Nikoloska$^{\star}$}%
    , \IEEEauthorblockN{Wim van Houtum$^{\star\ddagger}$}%
    , \IEEEauthorblockN{Yan Wu$^{\ddagger}$}%
    , and \IEEEauthorblockN{Alex Alvarado$^{\star}$}%
    \\
    \IEEEauthorblockA{\textit{$^{\star}$Information and Communication Theory Lab, Eindhoven University of Technology, The Netherlands}}\\
    \IEEEauthorblockA{\textit{$^{\dagger}$Communications Engineering Lab, Karlsruhe Institute of Technology, Germany}}\\
    \IEEEauthorblockA{\textit{$^{\ddagger}$NXP Semiconductors, Eindhoven, The Netherlands}}\\
    \IEEEauthorblockA{c.h.chen@tue.nl}
}

\maketitle
\thispagestyle{empty}

\begin{abstract}
    \input{tex/0-abs}
\end{abstract}


\section{Introduction} \label{sec:intro}
\input{tex/1-intro}

\section{System Model} \label{sec:sys_model}
\input{tex/2-sys}

\section{HMM, BCJR, and the Baum-Welch algorithm}\label{sec:hmm}
\input{tex/3-hmm}

\section{Receiver Design} \label{sec:rec_design}
This section discusses the receiver design for a joint turbo-BW-equalization system, as depicted in Fig.~\ref{fig:block_rx}. First, we will introduce a reduced state BW channel estimator based on a modified trellis structure. Then, we will formulate the turbo equalization and show how the extrinsic information of the prior symbol is computed and passed to the modified BW channel estimator.

\subsection{Modified Baum-Welch Channel Estimator} \label{sec:modifedBW}
\input{tex/4-1-jointem}

\subsection{Turbo Equalization} \label{sec:turboeq}
\input{tex/4-2-turboeq}
\section{Numerical Simulations}\label{sec:results}
\input{tex/5-Simu}

\section{Conclusions}\label{sec:conc}
\input{tex/6-conc}

\section*{Acknowledgments}
\input{tex/ACKs}

\vspace{12pt}
\end{document}

%% file: tex/0-abs.tex
Blind estimation of intersymbol interference channels based on the Baum-Welch (BW) algorithm, a specific implementation of the expectation-maximization (EM) algorithm for training hidden Markov models, is robust and does not require labeled data. However, it is known for its extensive computation cost, slow convergence, and frequently converges to a local maximum. In this paper, we modified the trellis structure of the BW algorithm by associating the channel parameters with two consecutive states. This modification enables us to reduce the number of required states by half while maintaining the same performance. Moreover, to improve the convergence rate and the estimation performance, we construct a joint turbo-BW-equalization system by exploiting the extrinsic information produced by the turbo decoder to refine the BW-based estimator at each EM iteration. Our experiments demonstrate that the joint system achieves convergence in $10$~EM iterations, which is $8$~iterations less than a separate system design for a signal-to-noise ratio (SNR) of $4$~dB. Additionally, the joint system provides improved estimation accuracy with a mean square error (MSE) of $10^{-4}$ for an SNR of $6$~dB. We also identify scenarios where a joint design is not preferable, especially when the channel is noisy (e.g., SNR~$=2$~dB) and the decoder cannot provide reliable extrinsic information for a BW-based estimator.

%% file: tex/1-intro.tex
This paper studies the fundamental problem of channel estimation and equalization of communication over intersymbol interference (ISI) and additive white Gaussian noise (AWGN) environments. Conventional methods such as Minimum Mean-Square Error estimation (MMSE) and decision-feedback equalization (DFE) \cite{Proakis} have been well-established research topics. More recently, neural network-aided system design has drawn a lot of attention \cite{Shlezinger20, Shlezinger22, CHC24}, where the channel likelihood function is directly estimated from labeled training data. However, the above methods require additional pilot/training data, decreasing system throughput. To enhance data efficiency, blind channel estimation, performed directly from the transmitted data, has received significant attention over the past few decades. 

The Baum-Welch (BW) algorithm, a specific expectation maximization (EM) algorithm for a hidden Markov model (HMM), was initially proposed by \cite{Baum_72} and is a popular method for blind channel estimation. The BW-based estimator enables us to calculate the necessary likelihoods using Bayesian inference directly from the received data. In addition, state-of-the-art joint blind channel estimation and symbol detection \cite{Kaleh94, Ghosh92, Anton97, Tong98, Lopes01, Lopes01_2, Schmid24, karanov24} has been shown to approach the optimal receiver with perfect channel state information. For example, \cite{Kaleh94} proposed blind channel identification using the EM algorithm, and based on the channel parameter estimates, it delivers optimal symbol-by-symbol decisions. In \cite{karanov24}, a BW-based estimator is utilized to optimize the likelihood function and transition probability jointly for the ISI channel with impulsive noise, achieving performance that approaches the optimal detector with perfect channel state information.  

Moreover, with the development of powerful iterative decoding algorithms, turbo equalization systems \cite{Douillard95, Tuchler02} have been extensively used to perform channel equalization and decoding by iteratively exchanging extrinsic information. Inspired by this idea, coding-aided blind channel estimation systems have been proposed using several coding schemes \cite{Ha00,  Garcia03, Otnes04, Gunther05, Zhao10}. In \cite{Ha00}, a blind BW-based channel estimator is utilized to directly estimate the channel coefficients. This estimator is combined with channel equalization and turbo decoding. Additionally, the prior symbol information provided by the outer turbo decoder is fed back to the BW-based estimator in each EM iteration. In \cite{Garcia03}, the authors combine the trellis representation of the encoder with the ISI channel. The BW algorithm is then designed to update the mean and variance of a Gaussian distribution. 

While the concept of joint blind channel estimation and turbo equalization is well-established, its operational advantages over a standalone estimator design have not been sufficiently explored for different channel conditions. In our work, we developed a blind turbo-BW-equalization receiver, similar to the method proposed in \cite{Garcia03} but with a single convolutional decoder. A major contribution of this paper is that we have identified the operational conditions for a joint system. Although a joint design generally outperforms a separate design, our study suggests that a standalone estimator is more robust in highly noisy channel conditions as the BW-based estimator cannot derive benefit from unreliable prior information provided by the turbo decoding block in a joint system.

Furthermore, this study addresses a less discussed topic: the trellis structure of the BW-based estimator. It is widely recognized that the BW-based algorithm is computationally expensive due to the necessity of calculating posterior beliefs through a forward-backward algorithm, with the complexity growing exponentially as the number of system states increases. Traditionally, the BW algorithm defines the channel model parameters to be optimized as directly associated with each state (e.g., \cite{Baum_72,HMM}). However, in this paper, we modify the trellis structure by associating the channel output parameters (e.g., mean and variance) with two consecutive states. This modification reduces the number of states by half without any loss in performance. A similar trellis structure has been implicitly constructed in \cite{Kaleh94,Garcia03,Ha00}. In this work, we explicitly formulate the equation for constructing the trellis and highlight the distinctions from the conventional BW algorithm.

This paper is structured as follows: Sec.~\ref{sec:sys_model} introduces the transmitter design and the channel model. Sec.~\ref{sec:hmm} details the structure of the HMM, efficient posterior probability inference, and conventional BW learning algorithm. In Sec.~\ref{sec:rec_design}, we illustrate the turbo equalization receiver design with the modified BW-based channel estimator. Sec.~\ref{sec:results} presents the numerical simulation setup and the performance of the considered joint turbo-BW-equalization system. Finally, Sec.~\ref{sec:conc} summarizes the paper with concluding remarks.

%% file: tex/2-sys.tex

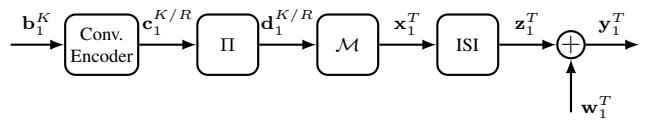
\begin{figure}[t]
    \centering
    {\input{tikz/block_TX_CH.tikz}} 
    \caption{System block of a convolutionally coded bit-interleaved BPSK transmitter over a linear ISI channel with AWGN.}   
    \label{fig:block_tx}
\end{figure}

The system model of transmission over a linear ISI channel with AWGN is shown in Fig.~\ref{fig:block_tx}. In the transmitter side, a serially concatenated encoder and a symbol mapper are constructed, where a convolutional encoder takes as input the information bit sequence $\mathbf{b}_1^K = (b_1, b_2, \cdots, b_{K})$ and generate coded bit sequence $\mathbf{c}_1^{K/R} = (c_1,c_2,\cdots,c_{K/R})$ with $ b_k, c_k \in \mathcal{U} = \{0,1\}$ based on a certain generator constraint with $L_{c}$ shift registers.\footnote{Throughout this paper, we use lowercase letters to represent a scalar with subscripts indicating a specific time instant and boldface letters to denote a sequence, with subscripts and superscripts denoting the start and end of the sequence. To keep the notation simple, we do not use a special notation for random variables.} Here, $R$ represents the coding rate. A bit-level interleaver ($\Pi$) is used to permute the convolutional encoder output, where the interleaved coded bit sequence is represented as $\mathbf{d}_1^{K/R}$. 
In this paper, we assume a binary PSK (BPSK) modulation, where a symbol mapper ($\mathcal{M}$) maps the binary coded bits to a symbol $x_t \in \mathcal{X}=\{+1,-1\}$. The symbol time step is denoted by the letter $t$ and the symbol sequence length $T = K/R$.

We consider the output sequence $\mathbf{y}_1^T$ of the channel to be affected by a linear ISI channel and additive Gaussian noise, $w_t$. The input-output relationship of the channel can be expressed as
\begin{align}\label{eq:channel_mod}
    y_t = \sum_{l=1}^L h_{l} x_{t-l+1}  + w_t,
\end{align}
where we use $z_t = \sum_{l=1}^L h_{l} x_{t-l+1}$ to represent the noiseless ISI channel output. The channel response $h_l$ is assumed to be time-invariant with $L$ denoting the channel memory. We assume the noise realizations $w_t \in \mathcal{N}(0,\sigma_w^2)$ to follow a Gaussian distribution with zero mean and variance $\sigma_w^2$. 

%% file: tikz/block_TX_CH.tikz
\begin{tikzpicture}[line width=1.5pt, 
block1/.style={rectangle, fill=white, text opacity=1, rounded corners, draw, inner sep=2pt, minimum width=8mm, minimum height=8mm, align=center, font=\scriptsize, thick},
linestyle/.style = {-latex,>={Latex[length=1mm,width=1mm]},thick},]

\tikzstyle{block2} = [block1, fill=orange!30]
\tikzstyle{block3} = [block1, fill=LimeGreen]

\def\dist{2.5}
\def\ttsize{\scriptsize}

\node[] at (0,\dist) (b) {};
\node[block1, right=0.3*\dist of b, align=center] (Inner) {Conv.\\Encoder};
\draw[linestyle] ([xshift=-20pt]Inner.west) -- (Inner.west) node[midway,above] {\ttsize{$\mathbf{b}_1^K$}};

\node[block1, right=0.3*\dist of Inner.east, align=center] (PI) {$\Pi$};
\draw[linestyle] (Inner.east) -- (PI.west) node[midway,above] {\ttsize{$\mathbf{c}_1^{K/R}$}};

\node[block1, right=0.3*\dist of PI.east, align=center] (Map) {$\mathcal{M}$};
\draw[linestyle] (PI.east) -- (Map.west) node[midway,above] {\ttsize{$\mathbf{d}_1^{K/R}$}};


\node[block1, right=0.3*\dist of Map.east, align=center] (ISI) {ISI};
\draw[linestyle] (Map.east) -- (ISI.west) node[midway,above] {\ttsize{$\mathbf{x}_1^T$}};

\node[draw, circle, inner sep=0pt, minimum size=0.3cm, thick, right=0.3*\dist of ISI.east] (AWGN) {$+$};
\draw[linestyle] ([yshift=-20pt]AWGN.south) -- (AWGN.south) node[pos=0.1, right] {\ttsize{$\mathbf{w}_1^T$}};

\draw[linestyle] (ISI.east) -- (AWGN.west) node[midway,above] {\ttsize{$\mathbf{z}_1^T$}};
\draw[linestyle] (AWGN.east) -- ([xshift=20pt]AWGN.east) node[midway, above] {\ttsize{$\mathbf{y}_1^T$}};


\end{tikzpicture}

%% file: tex/3-hmm.tex
In this section, we provide an overview of the structure of the HMM, its efficient inference algorithm for computing posterior probability, and the learning algorithm to update the model parameters. The inference algorithm will be used for maximum a-posteriori (MAP) decoding and equalization, and the learning algorithm will be utilized to design a BW-based estimator. These applications will be further explained in Sec.~\ref{sec:rec_design}. 

\subsection{Hidden Markov model}
An HMM is a statistical model that captures the temporal behavior of a sequence of states $\mathbf{s}_1^T$ from a Markov process via a sequence of state-dependent observations $\mathbf{y}_1^T$. Here, we denote $s_t$ as the actual state realization at time $t$, which is drawn from a time-invariant finite-state machine with a size $N$ alphabet $\mathcal{S}=\{S^0,\cdots,S^{N-1}\}$. The initial state distribution is defined as $p(s_1)$. The state-transition probability is stationary and can be written as $p(s_t = S^j | s_{t-1}=S^i)$ of moving from state $S^i$ at time $t-1$ to state $S^j$ at time $t$. The observation likelihood is represented as $p(y_t|s_t=S^j)$, which describes the probability of an observation $y_t$ being generated from a state $S^j$ at time $t$. Given the number of states $N$, an HMM can thus be fully described by a set of parameters
\begin{align}\label{eq:hmmpara}
    \lambda = \{p(s_1), p(s_t|s_{t-1}), p(y_t | s_t)\}.
\end{align}

\subsection{BCJR inference algorithm}
The Markovian structure of an HMM allows an efficient inference of the hidden states through a forward-backward algorithm. Specifically, the inference problem aims at calculating the posterior belief of a latent state at a specific time instant given the full sequence of observation as
\begin{equation} \label{eq:map}
    p(s_t | \mathbf{y}^T_1) \propto p(s_t, \mathbf{y}^T_1) = \sum_{s_{t-1}\in\mathcal{S}}{p(s_{t}, s_{t-1}, \mathbf{y}^T_1)}.
\end{equation}
We can further break up the joint probability in the right-hand side of~\eqref{eq:map} as
\begin{align} 
    p(s_t, &s_{t-1},\mathbf{y}^T_1) = p(s_{t},s_{t-1},\mathbf{y}_{t+1}^{T}, y_t, \mathbf{y}^{t-1}_1) \nonumber\\
    &= p(s_{t-1}, \mathbf{y}_1^{t-1}) \cdot p(y_t, s_t | s_{t-1}) \cdot p(\mathbf{y}_{t+1}^{T} | s_{t}) \label{eq:bcjr3}
\end{align}
using Bayes' rule and the Markovian assumption. Let $\alpha(s_{t-1})$, $\beta(s_t)$, and $\gamma(s_t,s_{t-1})$ denote  $p(s_{t-1}, \mathbf{y}_1^{t-1})$, $p(\mathbf{y}_{t+1}^{T} | s_{t})$, and $p(y_t, s_t | s_{t-1})$, respectively. These parameters are defined as the forward recursion, backward recursion, and the branch metric. 

Using the Bayes' rule, the branch metric can be further derived as 
\begin{align} \label{eq:gamma}
        \gamma(s_{t}, s_{t-1}) =  p(y_t| s_t) \cdot p(s_{t} | s_{t-1}), 
\end{align}
which can be seen as the multiplication of observation likelihood and the state transition probability.
The forward recursion representation of $\alpha$ can be derived as
\begin{align} 
    \alpha(s_{t}) &= p(s_{t}, \mathbf{y}_1^{t}) = \sum_{s_{t-1}\in\mathcal{S}}p(s_{t},s_{t-1}, y_t, \mathbf{y}_1^{t-1}) \label{eq:siso_alpha1} \\
    &= \sum_{s_{t-1}\in\mathcal{S}} p(y_t, s_t | s_{t-1}) \cdot p(s_{t-1}, \mathbf{y}_1^{t-1}) \label{eq:siso_alpha2} \\
    &= \sum_{s_{t-1}\in\mathcal{S}} \gamma(s_{t}, s_{t-1}) \cdot \alpha(s_{t-1}) \label{eq:siso_alpha3},
\end{align}
where from \eqref{eq:siso_alpha1} to \eqref{eq:siso_alpha2}, Bayes' rule and the Markovian property are used.
Following a similar approach, we can represent the backward recursion $\beta$ as
\begin{align} \label{eq:siso_beta}
    \beta(s_{t-1}) &= p(\mathbf{y}_{t}^{T} | s_{t-1}) =  \sum_{s_{t}\in\mathcal{S}}p(s_t, \mathbf{y}_{t+1}^{T}, y_t | s_{t-1}) \nonumber \\
    &= \sum_{s_{t}\in\mathcal{S}} p(\mathbf{y}_{t+1}^{T} | s_t) \cdot p(y_t,s_t | s_{t-1}) \nonumber \\
    &= \sum_{s_{t}\in\mathcal{S}} \beta(s_t) \cdot \gamma(s_{t}, s_{t-1}).
\end{align}

The calculations for the posterior beliefs $p(s_t | \mathbf{y}^T_1)$ using the forward-backward recursions outlined in \eqref{eq:map}--\eqref{eq:siso_beta} leads to the widely recognized Bahl-Cocke-Jelinek-Raviv (BCJR) algorithm \cite{bcjr74}.

\subsection{Baum-Welch learning algorithm}
We have discussed the general HMM structure and its efficient BCJR inference algorithm. In the following, we will explain how to use the unlabeled observation $\mathbf{y}_1^T$ to learn the HMM parameters $\lambda$ in \eqref{eq:hmmpara} using the well-known BW algorithm proposed in~\cite{Baum_72}, a specific application of the EM algorithm for HMMs. The BW learning procedure first computes the posterior beliefs $p(s_t|\mathbf{y}_1^T, \lambda)$ via the BCJR algorithm \eqref{eq:map}--\eqref{eq:siso_beta} with a fixed set of model parameters $\lambda$ \mbox{(E step)}. Then, the algorithm updates the model parameters $\lambda$ through a maximum likelihood reestimation procedure \mbox{(M step)}, thus completing one EM iteration. In this paper, we focus on the optimization of the likelihood function $p(y_t|s_t)$ and assume a fixed initial state probability $p(s_1)$ and state transition matrix $p(s_t|s_{t-1})$. For a comprehensive discussion on the optimization of the complete model parameter set $\lambda$ of an HMM, interested readers are referred to \cite{Baum_72, HMM}.

Considering a continuous observation $y_t$ that can be approximated by a Gaussian distribution, the BW algorithm estimates the likelihood function $p(y_t|s_t)$ via optimizing the means $\hat{\mu}_l$ and variances $\hat{\sigma}_l^2$ of the Gaussian distribution 
\begin{align}\label{eq:em_lik}
    p(y_t|s_t=S^l) = \frac{1}{\sqrt{2\pi \hat{\sigma}^2_l}}\exp{\left(-\frac{(y_t-\hat{\mu}_l)^2}{2\hat{\sigma}^2_l}\right)}.
\end{align}
The means of the observation likelihood function for state $S^j$ is updated through 
\begin{align}
    \hat{\mu}_l = \frac{\sum_{t=1}^{T} p(s_t=S^l|\mathbf{y}_1^T, \lambda) \cdot y_t} {\sum_{t=1}^{T} p(s_t=S^l|\mathbf{y}_1^T, \lambda)}.
\end{align}
Similarly, the update of the variances is performed as
\begin{align}\label{eq:HMM_sigma}
		\hat{\sigma}^2_l = \frac{\sum_{t=1}^{T} p(s_t=S^l|\mathbf{y}_1^T, \lambda) \cdot (y_t-\hat{\mu}_l)^2} {\sum_{t=1}^{T} p(s_t=S^l|\mathbf{y}_1^T, \lambda)}.
\end{align}
Based on the updated channel likelihood via \eqref{eq:em_lik}--\eqref{eq:HMM_sigma}, the BW algorithm proceeds to the next iteration by using the updated model parameters set $\hat{\lambda}$ to compute the posterior belief $p(s_t|\mathbf{y}_1^T, \hat{\lambda})$.

%% file: tex/4-1-jointem.tex
\begin{figure*}[t]
    \centering
    {\input{tikz/block_RX_Joint}} 
    \caption{Receiver block diagram of a joint turbo-BW-Equalization system design.}   
    \label{fig:block_rx}
\end{figure*}
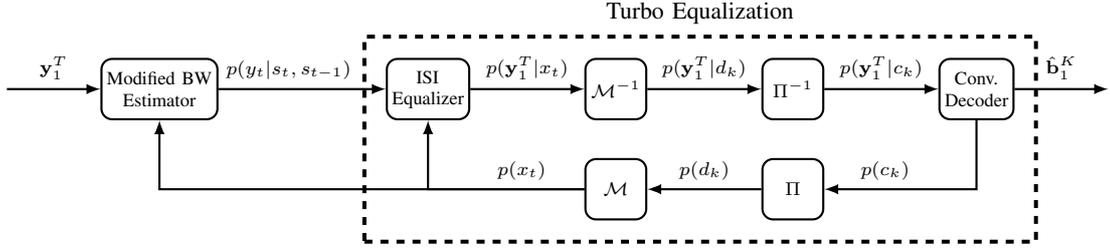
Conventionally, a BW-based estimator associates each set of channel parameters $\phi_l = \{\hat{\mu}_l,\hat{\sigma}_l^2\}$ with a specific state $S^l \in \mathcal{S}$ as in \eqref{eq:em_lik}, where $l \in |\mathcal{S}| = |\mathcal{X}|^L$ indicates the size of the channel parameters is the same as the trellis states. In this study, we implement a reduced state BW algorithm, where a new set of states $\dot{S}^j \in \dot{\mathcal{S}}$ with size $|\dot{\mathcal{S}}| = |\mathcal{X}|^{L-1}$ is defined and the channel parameters $\phi_l$ are associated with two consecutive states $s_t=\dot{S}^j$ and $s_{t-1}=\dot{S}^i$, namely the edges of the trellis. The likelihood function is then represented as 
\begin{align}\label{eq:lik_modem}
    p(y_t|s_t = \dot{S}^j,s_{t-1} = \dot{S}^i) = \frac{1}{\sqrt{2\pi \hat{\sigma}^2_l}}\exp{\left(-\frac{(y_t-\hat{\mu}_l)^2}{2\hat{\sigma}^2_l}\right)}.
\end{align}
The state transition probability of the underlying ISI channel shown in \eqref{eq:channel_mod} is represented as
\begin{align} \label{eq:state_tran}
    p(s_t={\dot{S}}^j|s_{t-1}={\dot{S}}^i) = 
    \begin{cases}
        p(x_t), & ({\dot{S}}^j, {\dot{S}}^i) \in {\dot{\mathcal{S}}}^2_{x} \\
        0, & \text{otherwise},
    \end{cases}
\end{align}
where $\dot{\mathcal{S}}^2_x$ denotes the set of state pairs $(\dot{S}^j, \dot{S}^i)$ that is driven by the symbol $x_t$.\footnote{Throughout this paper, we use calligraphic letters with superscript $2$ to denote the set of certain state pairs driven by the input symbol denoted by the subscript.} The trellis representation of the conventional BW algorithm and the reduced state implementation are shown in Figs.~\ref{fig:trellis1} and~\ref{fig:trellis2}, respectively. 

In a joint turbo-BW-equalization system design, the prior symbol probability $p(x_t$) in the right-hand-side of \eqref{eq:state_tran} is updated via the extrinsic information produced by the turbo equalization system, which will be illustrated shortly in the following subsection.

%% file: tikz/block_RX_Joint.tex
\begin{tikzpicture}[line width=1.5pt, 
block1/.style={rectangle, fill=white, text opacity=1, rounded corners, draw, inner sep=2pt, minimum width=8mm, minimum height=8mm, align=center, font=\scriptsize, thick},
linestyle/.style = {-latex,>={Latex[length=1mm,width=1mm]},thick},]

\tikzstyle{block2} = [block1, fill=gray!40]
\tikzstyle{block3} = [block1, fill=orange!30]

\def\dist{2.5}
\def\ttsize{\scriptsize}

\node[] at (0,\dist) (b) {};
\node[block1, right=0.6*\dist of b, align=center] (HMM) {Modified BW\\Estimator};
\draw[linestyle] ([xshift=-35pt]HMM.west) -- (HMM.west) node[midway,above] {\ttsize{$\mathbf{y}_1^T$}};

\node[block1, right=0.88*\dist of HMM.east, align=center] (det) {ISI\\Equalizer};
\draw[linestyle] (HMM.east) -- (det.west) node[pos=0.42, above] {\ttsize{$p({y_t} | s_t, s_{t-1})$}};

\node[block1, right=0.6*\dist of det.east, align=center] (demap) {$\mathcal{M}^{-1}$};
\draw[linestyle] (det.east) -- (demap.west) node[midway,above] {\ttsize{$p(\mathbf{y}_1^T | x_t)$}};

\node[block1, right=0.6*\dist of demap.east, align=center] (PIInv) {\ttsize{$\Pi^{-1}$}};

\draw[linestyle] (demap.east) -- (PIInv.west) node[midway,above] {\ttsize{$p(\mathbf{y}_1^T | d_k)$}};

\node[block1, right=0.6*\dist of PIInv.east, align=center] (CC) {Conv.\\Decoder};
\draw[linestyle] (PIInv.east) -- (CC.west) node[midway,above] {\ttsize{$p(\mathbf{y}_1^T | c_k)$}};
\draw[linestyle] (CC.east) -- ([xshift=35pt]CC.east) node[midway, above] {\ttsize{$\hat{\mathbf{b}}_1^K$}};

\node[block1, below=0.2*\dist of PIInv] (PI) {\ttsize{$\Pi$}};
\draw[linestyle] (CC.south) |- (PI.east) node[pos=0.8,above] {\ttsize{$p(c_k)$}};

\node[block1, below=0.2*\dist of demap] (map) {$\mathcal{M}$};
\draw[linestyle] (PI.west) -- (map.east) node[midway,above] {\ttsize{$p(d_k)$}};

\draw[linestyle] (map.west) -| (det.south) node[pos=0.2,above] {\ttsize{$p(x_t)$}};

\draw[linestyle] (map.west) -| (HMM.south);

\node[draw, dashed, fit=(det)(PI)(CC), inner sep=8pt, label=above: \small Turbo Equalization] (box) {};

\end{tikzpicture}

%% file: tex/4-2-turboeq.tex
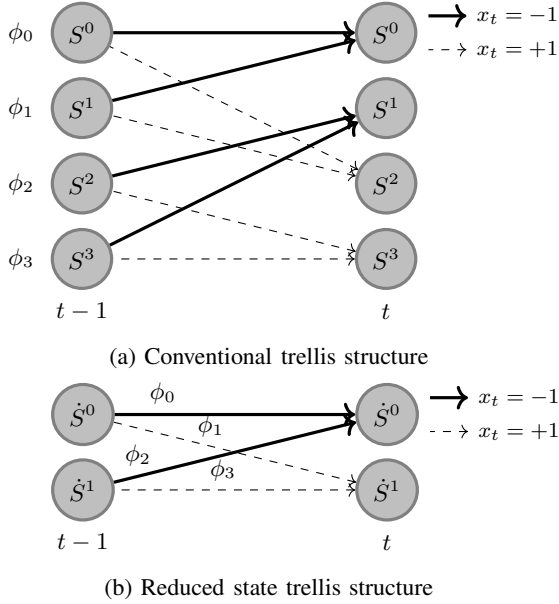
\begin{figure}[t]
    \centering
    \subfloat[Conventional trellis structure\label{fig:trellis1}]{%
        \centering
        \input{tikz/trellis_isi.tikz}
    }
    \vfill
    \subfloat[Reduced state trellis structure\label{fig:trellis2}]{%
        \centering
       {\input{tikz/trellis_isi2.tikz}} 
    }
    \caption{Trellis representations of BW-based channel estimators. $\phi_l = \{\hat{\mu}_l,\hat{\sigma}_l^2\}$ denotes the channel parameters set, which is associated with individual states in Fig.~\ref{fig:trellis1} and state pairs (edges) in Fig.~\ref{fig:trellis2}.}
\end{figure}

Given the state transition probability \eqref{eq:state_tran} and the updated likelihood function \eqref{eq:lik_modem}, the ISI equalizer calculates the posterior probability $p(s_t,s_{t-1},\mathbf{y}_1^T)$ using \eqref{eq:map}--\eqref{eq:siso_beta}. Then, the joint symbol probability can be computed via
\begin{align}
    p(x_t, \mathbf{y}_1^T) = \sum_{(\dot{S}^j, \dot{S}^i) \in \dot{\mathcal{S}}^2_{x}}{ {p(s_t=\dot{S}^j, s_{t-1}=\dot{S}^i, \mathbf{y}_1^T)} }. \nonumber
\end{align}

The turbo decoding principle involves passing only extrinsic information between the convolutional decoder and the ISI equalizer blocks. Extrinsic information refers to the information provided by one decoder (equalizer) about a specific bit (symbol), excluding the information used to derive it directly. In other words, it represents new information that can help the other decoder refine its estimates. We compute the extrinsic symbol information from the ISI equalizer as
\begin{align}
     p(\mathbf{y}_1^T|x_t) = \frac{p(x_t, \mathbf{y}_1^T)}{p(x_t)}. \nonumber
\end{align}
The prior probability of the input symbols is set to $p(x_t) = 1 / 2$ in the first turbo iteration.

To perform bit deinterleaving, we need to convert the symbol probability to bit probability as 
\begin{align}
    p(\mathbf{y}_1^T | d_k) = \mathcal{M}^{-1}[p(\mathbf{y}_1^T| x_t)]. \nonumber    
\end{align} 
After performing the bit-deinterleaving, the extrinsic coded bit information is 
\begin{align}
    p(\mathbf{y}_1^T|c_{k}) = \Pi^{-1} \left[ p( \mathbf{y}_1^T | d_k) \right],    
\end{align}
which is then fed to the convolutional decoder with a branch metric defined as
\begin{align}
    \gamma(s_k=\ddot{S}^j,s_{k-1}=\ddot{S}^i) &= p(s_k=\ddot{S}^j|s_{k-1}=\ddot{S}^i) \cdot p(\mathbf{y}_1^T|c_k) \nonumber \\
    & = p(b) \cdot p(\mathbf{y}_1^T|c_k), \quad (\ddot{S}^j, \ddot{S}^i) \in \ddot{\mathcal{S}}^2_{b}. \nonumber
\end{align}
Here, the convolutional decoder state is drawn from a state space $\ddot{\mathcal{S}} = \mathcal{U}^{L_{c}}$, where $|\ddot{\mathcal{S}}| = 2^{L_{c}}$.
Again, we can compute the joint probability $p(s_k,s_{k-1}, \mathbf{y}_1^T)$ via \eqref{eq:map}--\eqref{eq:siso_beta} and obtain the joint probability of the information bits as
\begin{align} 
    p(b_k, \mathbf{y}_1^T) =  {\sum_{(\ddot{S}^j,\ddot{S}^i) \in \ddot{\mathcal{S}}^2_{b}}{{ p(s_k = \ddot{S}^j,s_{k-1}=\ddot{S}^i, \mathbf{y}_1^T)} }}, \nonumber
\end{align}
and the joint probability of the coded bits as
\begin{align}
    p(c_k, \mathbf{y}_1^T) = \sum_{(\ddot{S}^j,\ddot{S}^i) \in \ddot{\mathcal{S}}^2_c} {{p(s_k=\ddot{S}^j,s_{k-1}=\ddot{S}^i, \mathbf{y}_1^T)}}. \nonumber
\end{align}
We can then compute the extrinsic information from the convolutional decoder as
\begin{align}
    p(c_k) = \frac{p(c_k,\mathbf{y}_1^T)}{p(\mathbf{y}_1^T|c_k)}.  \nonumber
\end{align}
After performing the bit interleaving as
\begin{align}
    p(d_k) = \Pi \left[ p(c_k) \right], \nonumber   
\end{align}
we can derive the extrinsic prior BPSK symbol information via a bit-to-symbol mapping
\begin{align}
    p(x_t) = \mathcal{M}[p(d_k)]. \nonumber
\end{align}
Finally, the updated prior symbol probability $p(x_t)$ is fed back to the ISI equalizer and the modified BW-based estimator in the next turbo iteration. Note that we refer to the turbo iteration as the entire turbo-BW-equalization loop. In contrast, the EM iteration refers to the optimization within the BW-based estimator.

%% file: tikz/trellis_isi.tikz
\tikzstyle{state}=[shape=circle,draw=black!50,fill=gray!50, font=\small]
\tikzstyle{observation}=[shape=rectangle,draw=orange!50,fill=orange!20]
\tikzstyle{lightedge}=[<-, dashed]
\tikzstyle{mainstate}=[state,very thick]
\tikzstyle{mainedge}=[<-,very thick]
\begin{tikzpicture}[]

\node  (s')         at (0,1.3)  {\small${t-1}$};
\node  (s1)         at (-.8,5) {\small$\phi_0$};
\node               at (-.8,4) {\small$\phi_1$};
\node               at (-.8,3) {\small$\phi_2$};
\node               at (-.8,2) {\small$\phi_3$};

\node[mainstate] (s1_1) at (0,5) {$S^0$};
\node[mainstate] (s2_1) at (0,4) {$S^1$};
\node[mainstate] (s3_1) at (0,3) {$S^2$};
\node[mainstate] (s4_1) at (0,2) {$S^3$};

\node               at (4,1.3) {\small$t$};
\node[mainstate] (s1_2) at (4,5) {$S^0$}
    edge[mainedge] (s1_1)
    edge[mainedge] (s2_1);
\node[mainstate] (s2_2) at (4,4) {$S^1$}
     edge[mainedge] (s3_1)
     edge[mainedge] (s4_1);

\node[mainstate] (s3_2) at (4,3) {$S^2$}
    edge[lightedge] (s1_1)
    edge[lightedge] (s2_1);
\node[mainstate] (s4_2) at (4,2) {$S^3$}
    edge[lightedge] (s3_1)
    edge[lightedge] (s4_1);

\matrix[
    anchor=west,
    fill=white,
    column sep = 0em
    ] at (s1_2.east)
    {
    \draw[mainedge]  (0.5, 0) -- (0,0); & \node[font=\footnotesize]  {$x_t=-1$}; \\
    \draw[lightedge] (0.5, 0) -- (0,0); & \node[font=\footnotesize] {$x_t=+1$}; \\
    };
\end{tikzpicture}

%% file: tikz/trellis_isi2.tikz
\tikzstyle{state}=[shape=circle,draw=black!50,fill=gray!50, font=\small]
\tikzstyle{observation}=[shape=rectangle,draw=orange!50,fill=orange!20]
\tikzstyle{lightedge}=[<-, dashed]
\tikzstyle{mainstate}=[state,very thick]
\tikzstyle{mainedge}=[<-,very thick]
\begin{tikzpicture}[]

\node  (s')         at (0, 3.3)  {\small${t-1}$};
\node  (s1)         at (-1,5) {\small};
\node               at (-1,4) {\small};

\node[mainstate] (s1_1) at (0,5) {$\dot{S}^0$};
\node[mainstate] (s2_1) at (0,4) {$\dot{S}^1$};

\node               at (4, 3.3) {\small$t$};
\node[mainstate] (s1_2) at (4,5) {$\dot{S}^0$}
    edge[mainedge] node[pos=0.8,above] {\small$\phi_0$} (s1_1) 
    edge[mainedge] node[pos=0.9,above] {\small$\phi_2$} (s2_1);
\node[mainstate] (s2_2) at (4,4) {$\dot{S}^1$}
     edge[lightedge] node[pos=0.6,above] {\small$\phi_1$} (s1_1)
     edge[lightedge] node[pos=0.55,above] {\small$\phi_3$} (s2_1);

\matrix[
    anchor=west,
    fill=white,
    column sep = 0em
    ] at (s1_2.east)
    {
    \draw[mainedge]  (0.5, 0) -- (0,0); & \node[font=\footnotesize]  {$x_t=-1$}; \\
    \draw[lightedge] (0.5, 0) -- (0,0); & \node[font=\footnotesize] {$x_t=+1$}; \\
    };
\end{tikzpicture}

%% file: tex/5-simu.tex
This section evaluates the performance of a blind turbo-BW-equalization system under a linear ISI channel with AWGN with the input-output relationship given in \eqref{eq:channel_mod}. A half-rate convolutional code with generators $(5,7)_8$, featuring a shift register with a length of $L_c=2$ is used. The number of information bits per data frame is $10000$. We terminate the trellis of the convolutional code by zero padding $2$~bits in each data frame, which leads to a bit interleaver with a $20004$~bits depth. The interleaved coded bits are then mapped to BPSK symbols. For all numerical experiments, we assume that the channel has a memory of length $L=3$~with a set of time-invariant coefficients $\mathbf{h}=(0.5, 0.71, 0.5)^T$, which is known to have the worst minimum Euclidean distance at its output according to \cite{Proakis}. Therefore, we have $|\mathcal{X}|^{L-1}=4$~states with $|\mathcal{X}|^{L}=8$~edges based on the modified trellis structure discussed in Sec.~\ref{sec:modifedBW}. We define the SNR at receiver as $\text{SNR}=||\mathbf{h}||^2 E\{|x_t|^2\}/\sigma_w^2 $.

This paper assumes that the receivers have perfect knowledge of the noise variance $\sigma^2_w$, as it is generally easier to estimate compared to the ISI channel coefficients in wireless communication systems. We set the variance associated with each trellis edge $l$ as $\hat{\sigma}^2_l = \sigma^2_w$. Instead, our focus is on evaluating the performance of the BW estimator in terms of optimizing the means $\hat{\mu}_l$ associated with each trellis edge. All systems are initialized with $\hat{\mu}_{l}  = \mu_l + \epsilon$, where~$\mu_l \in \left\{ \mathbf{h}^T \mathbf{x}^L \mid \mathbf{x}^L \in \mathcal{X}^L \right\}$ represents the perfect Gaussian mean and $\epsilon \in \mathcal{N}(0,0.1)$ is a Gaussian-distributed uncertainty with zero mean and variance $0.1$. All numerical results are averaged over $5000$ independent Monte-Carlo simulations. In contrast to the fixed prior symbol information $p(x_t)=1/2$ of a standalone BW estimator, the joint system updates $p(x_t)$ for each EM iteration in the BW-based channel estimator.

\begin{figure}[t]
    \centering
    {\input{tikz/MSE_IT.tikz}} 
    \caption{MSE of the estimated mean concerning the number of iterations for SNR $= 2$~dB (dotted line), $4$~dB (dashed line), and $6$~dB (solid line) for separate (black) and joint (orange) designs. Gray lines indicate the BW estimator with perfect extrinsic prior information feedback in \eqref{eq:geniepri}.} 
    \label{fig:mse_it}
\end{figure}
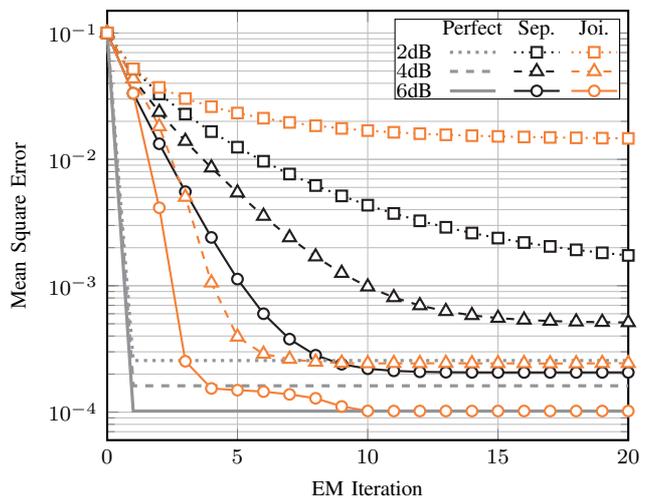

In Fig.~\ref{fig:mse_it}, we report the estimation performance for separate (standalone BW estimator and turbo equalization) and joint turbo-BW-equalization receiver designs. We evaluate the mean square error (MSE) of the estimated means $|| \mu_l - \hat{\mu}_l ||^2$ over EM iterations. A system with genie-aided perfect symbol information feedback, where we set
\begin{align}\label{eq:geniepri}
    p(x_t=1)=
    \begin{cases}
        1, & x_t = 1 \\
        0, & x_t = -1,
    \end{cases}
\end{align}
and $p(x_t=-1)=1-p(x_t=1)$~is reported as the gray lines in Fig.~\ref{fig:mse_it}. The results of the perfect feedback system converge to the correct estimates in just one EM iteration, showing that having accurate prior symbol extrinsic information is beneficial for the BW-based estimator. Additionally, the larger the SNR, the better the accuracy of the estimation and the faster the convergence rate for all system designs. The joint design shows significantly faster convergence compared to a separate design for an SNR of $4$~dB. Specifically, the joint system completes the convergence process approximately $8$~EM iterations sooner than a standalone BW estimator. This highlights the clear advantage of the combined approach in achieving rapid and efficient results. For SNR~$=6$~dB, the joint design can approach the optimal genie-added system in around $10$~EM iterations. Interestingly, a joint design fails to outperform the separate system design when the SNR is relatively low ($2$~dB). This observation implies that in the presence of significant channel noise, the extrinsic information produced by the turbo decoder may not be reliable. As a result, the BW-based estimator fails to take advantage of the feedback loop.

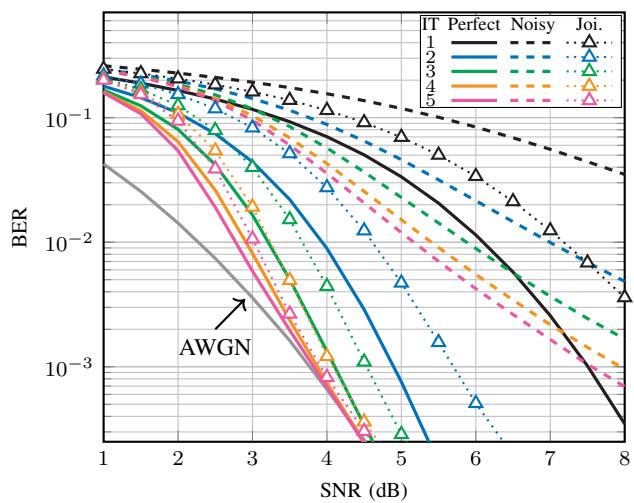
\begin{figure}[t]
    \centering
    {\input{tikz/BER_IT.tikz}} 
    \caption{BER performance of the joint turbo-BW-equalization system (dotted triangle) and turbo receiver with perfect (solid line) and noisy (dashed line) channel estimates. The gray solid line shows the optimal performance under an AWGN channel.}   
    \label{fig:ber_it}
\end{figure}

To evaluate receiver performance, we show the bit error rate (BER) for the joint turbo-BW-equalization system in Fig.~\ref{fig:ber_it}. As a reference, a turbo-equalization system that directly uses perfect channel estimates $\mu_l$ (solid line) and noisy channel estimates $\hat{\mu}_l$ (dashed line) are also shown. As expected, the noisy channel estimates significantly deteriorate the system performance. Incorporating a modified BW-based estimator already achieves a large gain in the first turbo iteration. By exploiting the feedback from the turbo equalization, the joint turbo-BW-equalization system approaches the performance of a system with perfect channel estimates after $4$~turbo iterations for SNRs greater than $4$~dB.

%% file: tikz/MSE_IT.tikz
\begin{tikzpicture}
	
	\begin{semilogyaxis}[
		axis line style = thick,
		grid = both,
		name = p1,
		xmin = 0, xmax=20,
		ymin = 6e-5, ymax=0.15,
		font=\footnotesize,
		ylabel = Mean Square Error, xlabel = EM Iteration,
		legend style={
			font=\tiny,
			nodes={scale=1.0},
		},
		legend cell align={left},
		legend pos = south west,
		]
        \addplot[color=Black, mark=*, mark options={fill=white}, thick] 
             table[x expr=\thisrowno{0}, y expr=\thisrowno{1}] 
            {data/mse_snr6.txt};\label{sep6} 
		
        \addplot[color=Orange, mark=*, mark options={fill=white}, thick] 
            table[x expr=\thisrowno{0}, y expr=\thisrowno{2}] 
            {data/mse_snr6.txt};\label{joi6} 
            
        \addplot[color=Gray, very thick] 
            table[x expr=\thisrowno{0}, y expr=\thisrowno{3}] 
            {data/mse_snr6.txt};\label{per6}
        
        \addplot[color=Black, mark=triangle*, mark options={solid, fill=white},  mark size = 2.8, thick, dashed] 
             table[x expr=\thisrowno{0}, y expr=\thisrowno{1}] 
            {data/mse_snr4.txt};\label{sep4} 
		
        \addplot[color=Orange,mark=triangle*, mark options={solid, fill=white},  mark size = 2.8, thick, dashed] 
            table[x expr=\thisrowno{0}, y expr=\thisrowno{2}] 
            {data/mse_snr4.txt};\label{joi4} 
            
        \addplot[color=Gray, very thick, dashed] 
            table[x expr=\thisrowno{0}, y expr=\thisrowno{3}] 
            {data/mse_snr4.txt};\label{per4}

        \addplot[color=Black, mark=square*, mark options={solid, fill=white},  mark size = 2, thick, dotted] 
             table[x expr=\thisrowno{0}, y expr=\thisrowno{1}] 
            {data/mse_snr2.txt};\label{sep2} 
		
        \addplot[color=Orange, mark=square*, mark options={solid, fill=white},  mark size = 2, thick, dotted] 
            table[x expr=\thisrowno{0}, y expr=\thisrowno{2}] 
            {data/mse_snr2.txt};\label{joi2} 
            
        \addplot[color=Gray, very thick, dotted] 
            table[x expr=\thisrowno{0}, y expr=\thisrowno{3}] 
            {data/mse_snr2.txt};\label{per2}

	\end{semilogyaxis}
 
\matrix[
    matrix of nodes,
    anchor = north east,
    fill = white, draw,
    inner sep = 0.1mm,
    column sep = 0.1mm,
    row sep = 0.4mm,
    node font=\footnotesize,
    column 1/.style={nodes={align=center}},
    column 2/.style={nodes={align=center}},
    column 3/.style={nodes={align=center}},
    column 4/.style={nodes={align=center}}
  ]
  at ([xshift=-3pt, yshift=-3pt]current axis.north east){
        & Perfect   & Sep. & Joi.\\ 
    $2$dB & \ref{per2} & \ref{sep2} & \ref{joi2}\\
    $4$dB & \ref{per4} & \ref{sep4} & \ref{joi4}\\
    $6$dB & \ref{per6} & \ref{sep6} & \ref{joi6}\\};

\end{tikzpicture}    

%% file: tikz/BER_IT.tikz
\begin{tikzpicture}
	
	\begin{semilogyaxis}[
		axis line style = thick,
		grid = both,
		name = p1,
		xmin = 1, xmax=8,
		ymin = 2.5e-4, ymax=0.7,
		font=\footnotesize,
		ylabel = BER, xlabel = SNR (dB),
		legend style={
			font=\tiny,
			nodes={scale=1.0},
		},
		legend cell align={left},
		legend pos = south west,
		]
        \addplot[color=Gray, very thick] 
             table[x expr=\thisrowno{0}, y expr=\thisrowno{1}] 
            {data/ber_awgn.txt};\label{awgn} 
            \draw[<-, thick] 
                (axis cs: 2.9, 3.5*10^-3) 
                -- (axis cs: 2.5, 2e-3); 
            \node at (axis cs: 2.5, 2*10^-3) [below] {\small{AWGN}};
            
        \addplot[color=Black, very thick] 
             table[x expr=\thisrowno{0}, y expr=\thisrowno{1}] 
            {data/ber_model_perfect.txt};\label{bper0} 
		
        \addplot[color=RoyalBlue, very thick] 
             table[x expr=\thisrowno{0}, y expr=\thisrowno{2}] 
            {data/ber_model_perfect.txt};\label{bper1} 

        \addplot[color=Green, very thick] 
             table[x expr=\thisrowno{0}, y expr=\thisrowno{3}] 
            {data/ber_model_perfect.txt};\label{bper2} 

        \addplot[color=BurntOrange, very thick] 
             table[x expr=\thisrowno{0}, y expr=\thisrowno{4}] 
            {data/ber_model_perfect.txt};\label{bper3} 
            
        \addplot[color=VioletRed, very thick] 
             table[x expr=\thisrowno{0}, y expr=\thisrowno{5}] 
            {data/ber_model_perfect.txt};\label{bper4} 

        \addplot[color=Black, dashed, very thick] 
             table[x expr=\thisrowno{0}, y expr=\thisrowno{1}] 
            {data/ber_model_var01.txt};\label{unc0} 
		
        \addplot[color=RoyalBlue, dashed, very thick] 
             table[x expr=\thisrowno{0}, y expr=\thisrowno{2}] 
            {data/ber_model_var01.txt};\label{unc1} 

        \addplot[color=Green, dashed, very thick] 
             table[x expr=\thisrowno{0}, y expr=\thisrowno{3}] 
            {data/ber_model_var01.txt};\label{unc2} 

        \addplot[color=BurntOrange, dashed, very thick] 
             table[x expr=\thisrowno{0}, y expr=\thisrowno{4}] 
            {data/ber_model_var01.txt};\label{unc3} 
            
        \addplot[color=VioletRed, dashed, very thick] 
             table[x expr=\thisrowno{0}, y expr=\thisrowno{5}] 
            {data/ber_model_var01.txt};\label{unc4} 
        \addplot[color=Black,  dotted, thick, mark=triangle*, mark options={ solid, fill = white}, mark size=2.8] 
             table[x expr=\thisrowno{0}, y expr=\thisrowno{1}] 
            {data/ber_em_joi_var01.txt};\label{em0} 
		
        \addplot[color=RoyalBlue,  dotted, thick, mark=triangle*, mark options={solid, fill = white}, mark size=2.8] 
             table[x expr=\thisrowno{0}, y expr=\thisrowno{2}] 
            {data/ber_em_joi_var01.txt};\label{em1} 

        \addplot[color=Green, dotted, thick, mark=triangle*, mark options={ solid, fill = white}, mark size=2.8] 
             table[x expr=\thisrowno{0}, y expr=\thisrowno{3}] 
            {data/ber_em_joi_var01.txt};\label{em2} 

        \addplot[color=BurntOrange, dotted, thick, mark=triangle*, mark options={solid, fill = white}, mark size=2.8] 
             table[x expr=\thisrowno{0}, y expr=\thisrowno{4}] 
            {data/ber_em_joi_var01.txt};\label{em3} 
            
        \addplot[color=VioletRed,  dotted, thick, mark=triangle*, mark options={solid, fill = white}, mark size=2.8] 
             table[x expr=\thisrowno{0}, y expr=\thisrowno{5}] 
            {data/ber_em_joi_var01.txt};\label{em4} 

		


            
	\end{semilogyaxis}
 
\matrix[
    matrix of nodes,
    anchor=north east,
    fill = white, draw,
    inner sep = 0.04em,
    column sep = 0em,
    node font=\scriptsize,
    column 1/.style={nodes={align=center}},
    column 2/.style={nodes={align=center}},
    column 3/.style={nodes={align=center}},
    column 4/.style={nodes={align=center}}
  ]
  at ([xshift=-1pt, yshift=-1pt]current axis.north east){
    IT  & Perfect   & Noisy & Joi. \\ 
    $1$ & \ref{bper0} & \ref{unc0} & \ref{em0} \\
    $2$ & \ref{bper1} & \ref{unc1} & \ref{em1} \\
    $3$ & \ref{bper2} & \ref{unc2} & \ref{em2} \\
    $4$ & \ref{bper3} & \ref{unc3} & \ref{em3} \\
    $5$ & \ref{bper4} & \ref{unc4} & \ref{em4} \\};

\end{tikzpicture}    

%% file: tex/6-conc.tex
This paper investigates the potential of enhancing blind channel estimation based on the BW algorithm over an ISI channel with AWGN. A modified BW-based estimator with a reconstructed trellis is implemented to reduce the total number of states by half compared to a traditional BW algorithm. Moreover, we employ a joint turbo-BW-equalization system, where the extrinsic information produced by the turbo system is fed back as the prior symbol information of the BW-based estimator. Our results identify the operational region of a joint system compared to a standalone BW estimator. The joint turbo-BW-equalization system generally outperforms a standalone BW estimator in terms of convergence rate and estimation accuracy. For an SNR of $4$~dB, the joint design converges in around $10$~EM iterations. In comparison, the separate system requires approximately $2$ times more EM iterations and converges to a worse MSE. However, for a noisy channel of an SNR of $2$~dB, A joint design cannot benefit from the feedback provided by a turbo decoder and thus performs worse than a standalone BW estimator. 

The preliminary results from this paper indicate that the quality of prior symbol probability significantly impacts the BW-based estimator. However, quantifying the reliability of extrinsic information regarding different SNR and initialization errors requires further investigation. Such metrics will serve as an important reference for determining the scheduling between the EM and the turbo iterations. In other words, they indicate how often we need to update the prior symbol information for the BW-based estimator. This study assumes perfect knowledge of the number of channel states, their variances, and the state transition probabilities, as well as a relatively small initialization error. Developing fully unsupervised and robust blind turbo receiver solutions remains a promising area for future research.

%% file: tex/ACKs.tex
This work was funded by the RAISE collaboration framework between Eindhoven University of Technology and NXP, including a PPS-supplement from the Dutch Ministry of Economic Affairs and Climate Policy.

%% file: main.bbl
\begin{thebibliography}{23}
\bibliographystyle{IEEEtran}

\bibitem{Proakis}
J. Proakis, ``Digital Communications, 3rd ed.'' \textit{NewYork: McGraw-Hill}, 1995.

\bibitem{Shlezinger20}
N. Shlezinger, Y. C. Eldar, N. Farsad, and A. J. Goldsmith, ``ViterbiNet: Symbol detection using a deep learning based Viterbi algorithm,'' \textit{IEEE 20th International Workshop on Signal Processing Advances in Wireless Communications (SPAWC)}, 2019.

\bibitem{Shlezinger22}
N. Shlezinger, N. Farsad, Y. C. Eldar, and A. J. Goldsmith, ``Learned factor graphs for inference from stationary time sequences,'' \textit{IEEE Trans. Signal Process.}, vol. 70, 2022.

\bibitem{CHC24}
C.-H. Chen, B. Karanov, W. van Houtum, W. Yan, A. Young and A. Alvarado, ``On the robustness of deep learning-aided symbol detectors to varying conditions and imperfect channel knowledge,'' in {\it{Proc. IEEE Wireless Communications and Networking Conference (WCNC)}}, 2024, pp. 1-6.

\bibitem{Baum_72}
L. E. Baum, ``An inequality and associated maximization technique in statistical estimation for probabilistic functions of Markov processes,'' {\it{Inequalities III: Proceedings of the 3rd Symposium on Inequalities}}, pp. 1-8, 1972.

\bibitem{Ghosh92}
M. Ghosh and C. L. Weber, ``Maximum-likelihood blind equalization,'' \textit{Optical Engineering}, vol. 31, no. 6, pp. 1224–1228, 1992.

\bibitem{Kaleh94}
G. K. Kaleh and R. Vallet, ``Joint parameter estimation and symbol detection for linear or nonlinear unknown channels,'' \textit{IEEE Trans. Commun.}, vol. 42, no. 7, pp. 2406-2413, 1994.

\bibitem{Anton97}
C. Anton-Haro, J.A.R. Fonollosa, and J.R. Fonollosa, ``Blind channel estimation and data detection using hidden markov models,'' \textit{IEEE Trans. on Sig. Process.}, vol.45, no. 1, pp. 241–247, 1997.

\bibitem{Tong98}
L. Tong and S. Perreau, ``Multichannel blind identification: From subspace to maximum likelihood methods,'' \textit{Proc. IEEE}, vol. 86, no. 10, pp. 1951–1968, Oct. 1998.

\bibitem{Lopes01}
R. R. Lopes, J.R. Barry, ``Exploiting error-control coding in blind channel estimation,'' \textit{Proc. IEEE Global Telecommunications Conference (GLOBECOM)}, pp. 1317-1321, 2001.

\bibitem{Lopes01_2}
R. R. Lopes and J. R. Barry, ``Blind iterative channel identification and equalization,'' \textit{Proc. IEEE International Conference on Communications (ICC)}, pp. 2256-2260, 2001.

\bibitem{Schmid24}
L. Schmid, T. Raviv, N. Shlezinger, and Laurent Schmalen, ``Blind channel estimation and joint symbol detection with data-driven factor graphs,'' \textit{ArXiv preprint arXiv:2401.12627}, 2024.

\bibitem{karanov24}
B. Karanov, C.-H. Chen, Y. Wu, A. Young, and W. van Houtum, ``Data-driven symbol detection for intersymbol interference channels with bursty impulsive noise,'' \textit{ArXiv preprint arXiv:2405.10814}, 2024.

\bibitem{Niu05}
H. Niu, M. Shen, J. A. Ritcey, and H. Liu, ``A factor graph approach to iterative channel estimation and LDPC decoding over fading channels,''
\textit{IEEE Trans. Wireless Commun.}, vol. 4, no. 4, pp. 1345–1350, Jul. 2005.

\bibitem{Douillard95}
C. Douillard, M. Jezequel, C. Berrou, A. Picart, P. Didier, and A. Glavieux, ``Iterative correction of intersymbol interference: turbo equalization,'' \textit{Eur. Trans. Telecommun.}, pp. 507–511, Sep./Oct. 1995.

\bibitem{Tuchler02}
M. Tuchler, R. Koetter and A. Singer, ``Turbo equalization: principles and new results'', \textit{IEEE Trans. Commun.}, vol. 50, no. 5, pp. 754-767, 2002.

\bibitem{Ha00}
P. Ha and B. Honary, ``Improved blind turbo detector,'' \textit{Proc. IEEE Vehicular Technology Conference (VTC)}, vol. 2, pp. 1196–1199, 2000.

\bibitem{Garcia03}
J. Garcia-Frias and J. D. Villasenor, ``Combined turbo detection and decoding for unknown ISI channels,'' \textit{IEEE Trans. Commun.}, vol. 51,
no. 1, pp. 79–85, Jan. 2003

\bibitem{Otnes04}
R. Otnes and M. Tuchler, ``Iterative channel estimation for turbo equalization of time-varying frequency-selective channels,'' \textit{IEEE Trans. Wireless Commun.}, vol. 3, no. 6, pp. 1918–1923, Nov. 2004.

\bibitem{Gunther05}
J. H. Gunther, M. Ankapura, and T. K. Moon, ``A generalized LDPC decoder for blind turbo equalization,'' \textit{IEEE Trans. Signal Process.}, vol. 53, part 1, pp. 3847-3856, Oct. 2005.

\bibitem{Zhao10}
X. Zhao and M. Davies, ``Coding-assisted blind MIMO separation and decoding,'' \textit{IEEE Trans. Veh. Technol.}, vol. 59, no. 9, pp. 4408–4417, Nov. 2010.

\bibitem{HMM}
L. R. Rabiner, ``A tutorial on hidden Markov models and selected applications in speech recognition,'' \textit{Proc. IEEE}, vol. 77, no. 2, pp. 257-286, 1989.


\bibitem{bcjr74}
L. R. Bahl, J. Cocke, F. Jelinek, and J. Raviv, ``Optimal decoding of linear codes for minimizing symbol error rate,'' \textit{IEEE Trans. Inform. Theory}, vol. 20, no. 2, pp. 284–287, Mar. 1974.






\end{thebibliography}
